arXiv:cs/0612106v1 [cs.LO] 21 Dec 2006

# On completeness of logical relations for monadic types *

Sławomir Lasota[1] ** David Nowak[2] Yu Zhang[3] ***

[1] Institute of Informatics, Warsaw University, Warszawa, Poland
[2] Research Center for Information Security,
National Institute of Advanced Industrial Science and Technology, Tokyo, Japan
[3] Project Everest, INRIA Sophia-Antipolis, France

**Abstract.** Software security can be ensured by specifying and verifying security properties of software using formal methods with strong theoretical bases. In particular, programs can be modeled in the framework of lambda-calculi, and interesting properties can be expressed formally by contextual equivalence (a.k.a. observational equivalence). Furthermore, imperative features, which exist in most real-life software, can be nicely expressed in the so-called computational lambda-calculus. Contextual equivalence is difficult to prove directly, but we can often use logical relations as a tool to establish it in lambda-calculi. We have already defined logical relations for the computational lambda-calculus in previous work. We devote this paper to the study of their completeness w.r.t. contextual equivalence in the computational lambda-calculus.

## 1 Introduction

**Contextual equivalence.** Two programs are contextually equivalent (a.k.a. observationally equivalent) if they have the same observable behavior, i.e. an outsider cannot distinguish them. Interesting properties of programs can be expressed using the notion of contextual equivalence. For example, to prove that a program does not leak a secret, such as the secret key used by an ATM to communicate with the bank, it is sufficient to prove that if we change the secret, the observable behavior will not change [18,3,19]: whatever experiment a customer makes with the ATM, he or she cannot guess information about the secret key by observing the reaction of the ATM. Another example is to specify functional properties by contextual equivalence. For example, if $\mathsf{sorted}$ is a function which checks that a list is sorted and $\mathsf{sort}$ is a function which sorts a list, then, for all list $l$, you want the expression $\mathsf{sorted}(\mathsf{sort}(l))$ to be contextually equivalent to the expression $\mathsf{true}$. Finally, in the context of parameterized verification, contextual equivalence allows the verification for all instantiations of the parameter to be reduced to the

* Partially supported by the RNTL project Prouvé, the ACI Sécurité Informatique Rossignol, the ACI jeunes chercheurs “Sécurité informatique, protocoles cryptographiques et détection d’intrusions”, and the ACI Cryptologie “PSI-Robuste”.

** Partially supported by the Polish KBN grant No. 4 T11C 042 25 and by the European Community Research Training Network *Games*. This work was performed in part during the author’s stay at LSV.

*** This work was mainly done when the author was a PhD student under an MENRT grant on ACI Cryptologie funding, École Doctorale Sciences Pratiques (Cachan).

verification for a finite number of instantiations (See e.g. [6] where logical relations are one of the essential ingredients).

**Logical relations.** While contextual equivalence is difficult to prove directly because of the universal quantification over contexts, logical relations [15,8] are powerful tools that allow us to deduce contextual equivalence in typed $\lambda$-calculi. With the aid of the so-called Basic Lemma, one can easily prove that logical relations are sound w.r.t. contextual equivalence. However, completeness of logical relations is much more difficult to achieve: usually we can only show the completeness of logical relations for types up to first order.

On the other hand, the computational $\lambda$-calculus [10] has proved useful to define various notions of computations on top of the $\lambda$-calculus: partial computations, exceptions, state transformers, continuations and non-determinism in particular. Moggi's insight is based on categorical semantics: while categorical models of the standard $\lambda$-calculus are cartesian closed categories (CCCs), the computational $\lambda$-calculus requires CCCs with a strong monad. Logical relations for monadic types, which are particularly introduced in Moggi's language, can be derived by the construction defined in [2] where soundness of logical relations is guaranteed.

However, monadic types introduce new difficulties. In particular, contextual equivalence becomes subtler due to the different semantics of different monads: equivalent programs in one monad are not necessarily equivalent in another! This accordingly makes completeness of logical relations more difficult to achieve in the computational $\lambda$-calculus. In particular the usual proofs of completeness up to first order do not go through.

**Contributions.** We propose in this paper a notion of contextual equivalence for the computational $\lambda$-calculus. Logical relations for this language are defined according to the general derivation in [2]. We then explore the completeness and we prove that for the partial computation monad, the exception monad and the state transformer monad, logical relations are still complete up to first-order types. In the case of the non-determinism monad, we need to restrict ourselves to a subset of first-order types. As a corollary, we prove that strong bisimulation is complete w.r.t. contextual equivalence in a $\lambda$-calculus with monadic non-determinism.

Not like previous work on using logical relations to study contextual equivalence in models with computational effects [16,13,11], most of which focus on computations with local states, our work in this paper is based on a more general framework for describing computations, namely the computational $\lambda$-calculus. In particular, very different forms of computations like continuations and non-determinism are studied, not just those for local states.

**Plan.** The rest of this paper is structured as follows: we devote Section 2 to preliminaries, by introducing basic knowledge of logical relations in a simple version of typed $\lambda$-calculus; then from Section 3 on, we move to the computational $\lambda$-calculus and we rest on a set-theoretical model. In particular, Section 3.4 sketches out the proof scheme of completeness of logical relations for monadic types and shows the difficulty of getting a general proof; we then switch to case studies and we explore, in Section 4, the completeness in the computational $\lambda$-calculus for a list of common monads: partial computations, exceptions, state transformers, continuations and the non-determinism; the last section consists of a discussion on related work and perspectives.

## 2 Logical relations for the simply typed $\boldsymbol{\lambda}$-calculus

### 2.1 The simply typed $\boldsymbol{\lambda}$-calculus $\boldsymbol{\lambda^{\rightarrow}}$

Let $\lambda^{\rightarrow}$ be a simple version of typed $\lambda$-calculus:

$$\begin{array}{lll} \text{Types:} & \tau, \tau', \ldots ::= & b \mid \tau \rightarrow \tau' \\ \text{Terms:} & t, t', \ldots \;::= & x \mid c \mid \lambda x \cdot t \mid tt' \end{array}$$

where $b$ ranges over a set of base types (booleans, integers, etc.), $c$ over a set of constants and $x$ over a set of variables. We write $t[u/x]$ the result of substituting the term $u$ for free occurrences of the variable $x$ in the term $t$. Typing *judgments* are of the form $\mathit{\Gamma} \vdash t : \tau$ where $\mathit{\Gamma}$ is a *typing context*, i.e. a finite mapping from variables to types. We say that $x : \tau$ is in $\mathit{\Gamma}$ whenever $\mathit{\Gamma}(x) = \tau$. We write $\mathit{\Gamma}, x : \tau$ for the typing context which agrees with $\mathit{\Gamma}$ except that it maps $x$ to $\tau$. Typing rules are as standard. We consider the set theoretical semantics of $\lambda^{\rightarrow}$. The semantics of any type $\tau$ is given by a set $[\![\tau]\!]$. Those sets are such that $[\![\tau \rightarrow \tau']\!]$ is the set of all functions from $[\![\tau]\!]$ to $[\![\tau']\!]$, for all types $\tau$ and $\tau'$. A $\mathit{\Gamma}$*-environment* $\rho$ is a map such that, for every $x : \tau$ in $\mathit{\Gamma}$, $\rho(x)$ is an element of $[\![\tau]\!]$. We write $\rho[x := a]$ for the environment which agrees with $\rho$ except that it maps $x$ to $a$. We write $[x := a]$ for the environment just mapping $x$ to $a$. Let $t$ be a term such that $\mathit{\Gamma} \vdash t : \tau$ is derivable. The denotation of $t$, w.r.t. a $\mathit{\Gamma}$-environment $\rho$, is given as usual by an element $[\![t]\!]\rho$ of $[\![\tau]\!]$. We write $[\![t]\!]$ instead of $[\![t]\!]\rho$ when $\rho$ is irrelevant, e.g., when $t$ is a closed term. When given a value $a \in [\![\tau]\!]$, we say that it is *definable* if and only if there exists a closed term $t$ such that $\vdash t : \tau$ is derivable and $a = [\![t]\!]$.

Let $\mathbf{Obs}$ be a subset of base types, called *observation types*, such as booleans, integers, etc. A *context* $\mathbb{C}$ is a term such that $x : \tau \vdash \mathbb{C} : o$ is derivable, where $o$ is an observation type. We spell the standard notion of *contextual equivalence* in a denotational setting: two elements $a_1$ and $a_2$ of $[\![\tau]\!]$, are *contextually equivalent* (written as $a_1 \approx_\tau a_2$), if and only if for any context $\mathbb{C}$ such that $x : \tau \vdash \mathbb{C} : o$ ($o \in \mathbf{Obs}$) is derivable, $[\![\mathbb{C}]\!][x := a_1] = [\![\mathbb{C}]\!][x := a_2]$. We say that two closed terms $t_1$ and $t_2$ of the same type $\tau$ are *contextually equivalent* whenever $[\![t_1]\!] \approx_\tau [\![t_2]\!]$. Without making confusion, we shall use the same notation $\approx_\tau$ to denote the contextual equivalence between terms. We also define a relation $\sim_\tau$: for every pair of values $a_1, a_2 \in [\![\tau]\!]$, $a_1 \sim_\tau a_2$ if and only if $a_1, a_2$ are definable and $a_1 \approx_\tau a_2$.

### 2.2 Logical relations

Essentially, a (binary) *logical relation* [8] is a family $(\mathcal{R}_\tau)_{\tau \text{ type}}$ of relations, one for each type $\tau$, on $[\![\tau]\!]$ such that related functions map related arguments to related results. More formally, it is a family $(\mathcal{R}_\tau)_{\tau \text{ type}}$ of relations such that for every $f_1, f_2 \in$

$[\![\tau \to \tau']\!]$,

$$f_1 \ \mathcal{R}_{\tau\to\tau'} \ f_2 \quad \iff \quad \forall a_1, a_2 \in [\![\tau]\!] \ . \ a_1 \ \mathcal{R}_\tau \ a_2 \implies f_1(a_1) \ \mathcal{R}_{\tau'} \ f_2(a_2)$$

There is no constraint on relations at base types. In $\lambda^\to$, once the relations at base types are fixed, the above condition forces $(\mathcal{R}_\tau)_{\tau \text{ type}}$ to be uniquely determined by induction on types. We might have other complex types, e.g., products in variations of $\lambda^\to$, and in general, relations of these complex types should be also uniquely determined by relations of their type components. For instance, pairs are related when their elements are pairwise related. A unary logical relation is also called a *logical predicate*.

A so-called *Basic Lemma* comes along with logical relations since Plotkin's work [15]. It states that if $\Gamma \vdash t : \tau$ is derivable, $\rho_1$, $\rho_2$ are two related $\Gamma$-environments, and every constant is related to itself, then $[\![t]\!]\rho_1 \ \mathcal{R}_\tau \ [\![t]\!]\rho_2$. Here two $\Gamma$-environments $\rho_1$, $\rho_2$ are related by the logical relation, if and only if $\rho_1(x) \ \mathcal{R}_\tau \ \rho_2(x)$ for every $x : \tau$ in $\Gamma$. Basic Lemma is crucial for proving various properties using logical relations [8]. In the case of establishing contextual equivalence, it implies that, for every context $\mathbb{C}$ such that $x : \tau \vdash \mathbb{C} : o$ is derivable ($o \in \mathbf{Obs}$), $[\![\mathbb{C}]\!][x := a_1] \ \mathcal{R}_o \ [\![\mathbb{C}]\!][x := a_2]$ for every pair of related values $a_1, a_2$ in $[\![\tau]\!]$. If $\mathcal{R}_o$ is the equality, then $[\![\mathbb{C}]\!][x := a_1] = [\![\mathbb{C}]\!][x := a_2]$, i.e., $a_1 \approx_\tau a_2$. Briefly, for every logical relation $(\mathcal{R}_\tau)_{\tau \text{ type}}$ such that $\mathcal{R}_o$ is the equality for every observation type $o$, logically related values are necessarily contextually equivalent, i.e., $\mathcal{R}_\tau \subseteq \approx_\tau$ for any type $\tau$.

Completeness states the inverse: a logical relation $(\mathcal{R}_\tau)_{\tau \text{ type}}$ is *complete* if every contextually equivalent values are related by this logical relation, i.e., $\approx_\tau \subseteq \mathcal{R}_\tau$ for every type $\tau$. Completeness for logical relations is hard to achieve, even in a simple version of $\lambda$-calculus like $\lambda^\to$. Usually we are only able to prove completeness for types up to first order (the order of types is defined inductively: $\mathbf{ord}(b) = 0$ for any base type $b$; $\mathbf{ord}(\tau \to \tau') = \max(\mathbf{ord}(\tau) + 1, \mathbf{ord}(\tau'))$ for function types). The following proposition states the completeness of logical relations in $\lambda^\to$, for types up to first order:

**Proposition 1.** *There exists a logical relation $(\mathcal{R}_\tau)_{\tau \text{ type}}$ for $\lambda^\to$, with partial equality on observation types, such that if $\vdash t_1 : \tau$ and $\vdash t_2 : \tau$ are derivable, for any type $\tau$ up to first order, $t_1 \approx_\tau t_2 \implies [\![t_1]\!] \ \mathcal{R}_\tau \ [\![t_2]\!]$.*

*Proof.* Let $(\mathcal{R}_\tau)_{\tau \text{ type}}$ be the logical relation induced by $\mathcal{R}_b = \sim_b$ at every base type $b$ and we show that it is complete for types up to first order.

The proof is by induction over $\tau$. Case $\tau = b$ is obvious. Let $\tau = b \to \tau'$. Take two terms $t_1, t_2$ of type $b \to \tau'$ such that $[\![t_1]\!]$ and $[\![t_2]\!]$ are related by $\approx_{b\to\tau'}$. Let $f_1 = [\![t_1]\!]$ and $f_2 = [\![t_2]\!]$. Assume that $a_1, a_2 \in [\![b]\!]$ are related by $\mathcal{R}_b$, therefore $a_1 \sim_b a_2$ since $\mathcal{R}_b = \sim_b$. Clearly, $a_1$ and $a_2$ are thus definable, say by terms $u_1$ and $u_2$, respectively. Then, for any context $\mathbb{C}$ such that $x : \tau' \vdash \mathbb{C} : o$ ($o \in \mathbf{Obs}$) is derivable,

$$\begin{aligned}
&[\![\mathbb{C}]\!][x := f_1(a_1)] \\
&= [\![\mathbb{C}[xu_1/x]]\!][x := f_1] \quad \text{(since } a_1 = [\![u_1]\!]\text{)} \\
&= [\![\mathbb{C}[xu_1/x]]\!][x := f_2] \quad \text{(since } f_1 \approx_{b\to\tau'} f_2\text{)} \\
&= [\![\mathbb{C}]\!][x := f_2(a_1)] \\
&= [\![\mathbb{C}[t_2x/x]]\!][x := a_1] \quad \text{(since } f_2 = [\![t_2]\!]\text{)} \\
&= [\![\mathbb{C}[t_2x/x]]\!][x := a_2] \quad \text{(since } a_1 \approx_b a_2\text{)} \\
&= [\![\mathbb{C}]\!][x := f_2(a_2)].
\end{aligned}$$

Hence $f_1(a_1) \approx_{\tau'} f_2(a_2)$. Moreover, $f_1(a_1)$ and $f_2(a_2)$ are therefore definable by $t_1 u_1$ and $t_2 u_2$ respectively. By induction hypothesis, $f_1(a_1)\ \mathcal{R}_{\tau'}\ f_2(a_2)$. Because $a_1$ and $a_2$ are arbitrary, we conclude that $f_1\ \mathcal{R}_{b\to\tau'}\ f_2$. □

Note that an equivalent way to state completeness of logical relations is to say that there exists a logical relation $(\mathcal{R}_\tau)_{\tau \text{ type}}$ which is partial equality on observation types and such that, for all first-order types $\tau$, $\sim_\tau \subseteq \mathcal{R}_\tau$.

## 3 Logical relations for the computational λ-calculus

### 3.1 The computational λ-calculus $\lambda_{Comp}$

From the section on, our discussion is based on another language — Moggi's computational $\lambda$-calculus. Moggi defines this language so that one can express various forms of side effects (exceptions, non-determinism, etc.) in this general framework [10]. The computational $\lambda$-calculus, denoted by $\lambda_{Comp}$, extends $\lambda^{\to}$:

$$\begin{array}{lll} \text{Types:} & \tau, \tau', \ldots ::= & b \mid \tau \to \tau' \mid \mathsf{T}\tau \\ \text{Terms:} & t, t', \ldots ::= & x \mid c \mid \lambda x \cdot t \mid tt' \mid \mathtt{val}(t) \mid \mathtt{let}\ x \Leftarrow t\ \mathtt{in}\ t' \end{array}$$

An extra unary type constructor $\mathsf{T}$ is introduced in the computational $\lambda$-calculus: intuitively, a type $\mathsf{T}\tau$ is the type of computations of type $\tau$. We call $\mathsf{T}\tau$ a *monadic type* in the sequel. The two extra constructs $\mathtt{val}(t)$ and $\mathtt{let}\ x \Leftarrow t\ \mathtt{in}\ t'$ represent respectively the trivial computation and the sequential computation, with the typing rules:

$$\frac{\Gamma \vdash t : \tau}{\Gamma \vdash \mathtt{val}(t) : \mathsf{T}\tau} \qquad\qquad \frac{\Gamma \vdash t : \mathsf{T}\tau \quad \Gamma, x : \tau \vdash t' : \mathsf{T}\tau'}{\Gamma \vdash \mathtt{let}\ x \Leftarrow t\ \mathtt{in}\ t' : \mathsf{T}\tau'}$$

Note that the `let` construct here should not be confused with that in PCF: in $\lambda_{Comp}$, we bind the result of the term $t$ to the variable $x$, but they are not of the same type — $t$ must be a computation.

Moggi also builds a categorical model for the computational $\lambda$-calculus, using the notion of monads [10]. Whereas categorical models of simply typed $\lambda$-calculi such as $\lambda^{\to}$ are usually cartesian closed categories (CCCs), a model for $\lambda_{Comp}$ requires additionally a strong monad $(T, \eta, \mu, \mathrm{t})$ be defined over the CCC. Consequently, a monadic type is interpreted using the monad $T$: $[\![\mathsf{T}\tau]\!] = T[\![\tau]\!]$, and each term in $\lambda_{Comp}$ has a unique interpretation as a morphism in a CCC with the strong monad [10]. Semantics of the two additional constructs can be given in full generality in a categorical setting [10]: the denotations of `val` construct and `let` construct are defined by the follwoing composites respectively:

$$[\![\Gamma \vdash \mathtt{val}(t) : \mathsf{T}\tau]\!] : \quad [\![\Gamma]\!] \xrightarrow{[\![\Gamma \vdash t:\tau]\!]} [\![\tau]\!] \xrightarrow{\eta_{[\![\tau]\!]}} T[\![\tau]\!],$$

$$\begin{aligned}[\![\Gamma \vdash \mathtt{let}\ x \Leftarrow t_1\ \mathtt{in}\ t_2 : \mathsf{T}\tau']\!] : [\![\Gamma]\!] &\xrightarrow{\langle \mathrm{id}_{[\![\Gamma]\!]}, [\![\Gamma \vdash t_1 : \mathsf{T}\tau]\!]\rangle} [\![\Gamma]\!] \times T[\![\tau]\!] \xrightarrow{\mathrm{t}_{[\![\Gamma]\!],[\![\tau]\!]}} T[\![\Gamma]\!] \times [\![\tau]\!] \\ &\xrightarrow{T[\![\Gamma, x:\tau \vdash t_2 : \mathsf{T}\tau']\!]} TT[\![\tau']\!] \xrightarrow{\mu_{[\![\tau']\!]}} T[\![\tau']\!].\end{aligned}$$

In particular, the interpretation of terms in the computational $\lambda$-calculus must satisfy the following equations:

$$[\![\texttt{let}\ x \Leftarrow \texttt{val}(t_1)\ \texttt{in}\ t_2]\!]\rho = [\![t_2[t_1/x]]\!]\rho, \tag{1}$$

$$[\![\texttt{let}\ x_2 \Leftarrow (\texttt{let}\ x_1 \Leftarrow t_1\ \texttt{in}\ t_2)\ \texttt{in}\ t_3]\!]\rho = [\![\texttt{let}\ x_1 \Leftarrow t_1\ \texttt{in}\ \texttt{let}\ x_2 \Leftarrow t_2\ \texttt{in}\ t_3]\!]\rho, \tag{2}$$

$$[\![\texttt{let}\ x \Leftarrow t\ \texttt{in}\ \texttt{val}(x)]\!]\rho = [\![t]\!]\rho. \tag{3}$$

We shall focus on Moggi's monads defined over the category $\mathcal{S}et$ of sets and functions. Figure 1 lists the definitions of some concrete monads: partial computations, exceptions, state transformers, continuations and non-determinism. We shall write $\lambda_{Comp}^{PESCN}$ to refer to $\lambda_{Comp}$ where the monad is restricted to be one of these five monads.

| | |
|---|---|
| Partial computation: | $[\![\mathsf{T}\tau]\!] = [\![\tau]\!] \cup \{\bot\}$ |
| | $[\![\texttt{val}(t)]\!]\rho = [\![t]\!]\rho$ |
| | $[\![\texttt{let}\ x \Leftarrow t_1\ \texttt{in}\ t_2]\!]\rho = \begin{cases} [\![t_2]\!]\rho[x := [\![t_1]\!]\rho], & \text{if } [\![t_1]\!]\rho \neq \bot \\ \bot, & \text{if } [\![t_1]\!]\rho = \bot \end{cases}$ |
| Exception: | $[\![\mathsf{T}\tau]\!] = [\![\tau]\!] \cup E$ |
| | $[\![\texttt{val}(t)]\!]\rho = [\![t]\!]\rho$ |
| | $[\![\texttt{let}\ x \Leftarrow t_1\ \texttt{in}\ t_2]\!]\rho = \begin{cases} [\![t_2]\!]\rho[x := [\![t_1]\!]\rho], & \text{if } [\![t_1]\!]\rho \notin E \\ [\![t_1]\!]\rho, & \text{if } [\![t_1]\!]\rho \in E \end{cases}$ |
| State transformer: | $[\![\mathsf{T}\tau]\!] = ([\![\tau]\!] \times St)^{St}$ |
| | $[\![\texttt{val}(t)]\!]\rho = \underline{\lambda}\, s \cdot ([\![t]\!]\rho, s)$ |
| | $[\![\texttt{let}\ x \Leftarrow t_1\ \texttt{in}\ t_2]\!]\rho = \underline{\lambda}\, s \cdot ([\![t_2]\!]\rho[x := a_1])s_1,$ |
| | where $a_1 = \pi_1(([\![t_1]\!]\rho)s), s_1 = \pi_2(([\![t_2]\!]\rho)s)$ |
| Continuation: | $[\![\mathsf{T}\tau]\!] = R^{R^{[\![\tau]\!]}}$ |
| | $[\![\texttt{val}(t)]\!]\rho = \underline{\lambda}\, k^{[\![\tau]\!] \to R} \cdot k([\![t]\!]\rho)$ |
| | $[\![\texttt{let}\ x \Leftarrow t_1\ \texttt{in}\ t_2]\!]\rho = \underline{\lambda}\, k^{[\![\tau_2]\!] \to R} \cdot ([\![t_1]\!]\rho)k'$ |
| | where $k'$ is a function: $\underline{\lambda}\, v^{[\![\tau_1]\!]} \cdot ([\![t_2]\!]\rho[x := v])k$ |
| Non-determinism: | $[\![\mathsf{T}\tau]\!] = \mathbb{P}_{\mathrm{fin}}([\![\tau]\!])$ |
| | $[\![\texttt{val}(t)]\!]\rho = \{[\![t]\!]\rho\}$ |
| | $[\![\texttt{let}\ x \Leftarrow t_1\ \texttt{in}\ t_2]\!]\rho = \bigcup_{a \in [\![t_1]\!]\rho} [\![t_2]\!]\rho[x := a]$ |

**Fig. 1.** Concrete monads defined in $\mathcal{S}et$

The computational $\lambda$-calculus is strongly normalizing [1]. The reduction rules in $\lambda_{Comp}$ are called $\beta c$-reduction rules in [1], which, apart from standard $\beta$-reduction in the $\lambda$-calculus, contains especially the following two rules for computations:

$$\texttt{let}\ x \Leftarrow \texttt{val}(t_1)\ \texttt{in}\ t_2 \ \to_{\beta c}\ t_2[t_1/x], \tag{4}$$

$$\texttt{let}\ x_2 \Leftarrow (\texttt{let}\ x_1 \Leftarrow t_1\ \texttt{in}\ t_2)\ \texttt{in}\ t \ \to_{\beta c}\ \texttt{let}\ x_1 \Leftarrow t_1\ \texttt{in}\ (\texttt{let}\ x_2 \Leftarrow t_2\ \texttt{in}\ t). \tag{5}$$

With respect to the $\beta c$ rules, every term can be reduced to a term in the $\beta c$-normal form. Considering also the following $\eta$-equality rule for monadic types [1]:

$$\texttt{let}\ x \Leftarrow t\ \texttt{in}\ t'[\texttt{val}(x)/x'] =_\eta t'[t/x'], \tag{6}$$

we can write every term of a monadic type in the following $\beta c$-normal $\eta$-long form

$$\texttt{let}\ x_1 \Leftarrow d_1 u_{11} \cdots u_{1k_1}\ \texttt{in} \cdots \texttt{let}\ x_n \Leftarrow d_n u_{n1} \cdots u_{nk_n}\ \texttt{in}\ \texttt{val}(u),$$

where $n = 0, 1, 2, \ldots$, every $d_i$ ($1 \leq i \leq n$) is either a constant or a variable, $u$ and $u_{ij}$ ($1 \leq i \leq n, 1 \leq j \leq k_j$) are all $\beta c$-normal terms or $\beta c$-normal-$\eta$-long terms (of monadic types). In fact, the rules (4-6) just identify the equations (1-3) respectively.

**Lemma 1.** *For every term $t$ of type $\mathsf{T}\tau$ in $\lambda_{Comp}$, there exists a $\beta c$-normal-$\eta$-long term $t'$ such that $[\![t']\!]\rho = [\![t]\!]\rho$, for every valid interpretation $[\![_]\!]\rho$ (i.e., interpretations satisfying the equations (1-3)).*

*Proof.* Because the computational $\lambda$-calculus is strongly normalizing, we consider the $\beta c$-normal form of term $t$ and prove it by the structural induction on $t$.

– If $t$ is either a variable, a constant or an application, according to the equation (3):

$$[\![t]\!]\rho = [\![\texttt{let}\ x \Leftarrow t\ \texttt{in}\ \texttt{val}(x)]\!]\rho.$$

In particular, if $t$ is an application $t_1 t_1$, then $t_1$ must be either a variable or a constant since $t$ is $\beta c$-normal. Therefore, the term $\texttt{let}\ x \Leftarrow t\ \texttt{in}\ \texttt{val}(x)$ is in the $\beta c$-normal-$\eta$-long form.

– If $t$ is a trivial computation $\texttt{val}(t')$, by induction there is a $\beta c$-normal-$\eta$-long term $t''$ such that $[\![t']\!]\rho = [\![t'']\!]\rho$, for every valid $\rho$, then $[\![\texttt{val}(t')]\!]\rho = [\![\texttt{val}(t'')]\!]\rho$ as well.

– If $t$ is a sequential computation $\texttt{let}\ x \Leftarrow t_1\ \texttt{in}\ t_2$, since it is $\beta c$-normal, $t_1$ should not be any $\texttt{val}$ or $\texttt{let}$ term — $t_1$ must be of the form $du_1 \cdots u_n$ ($n = 0, 1, 2, \ldots$) with $d$ either a variable or a constant. By induction, there is a $\beta c$-normal-$\eta$-long term $t_2'$ such that $[\![t_2]\!]\rho = [\![t_2]\!]\rho$, for every valid $\rho$, then $[\![t]\!]\rho = [\![\texttt{let}\ x \Leftarrow t_1'\ \texttt{in}\ t_2']\!]\rho$ and the latter is in the $\beta c$-normal-$\eta$-long form. □

### 3.2 Contextual equivalence for $\lambda_{Comp}$

As argued in [3], the standard notion of contextual equivalence does not fit in the setting of the computational $\lambda$-calculus. In order to define contextual equivalence for $\lambda_{Comp}$, we have to consider contexts $\mathbb{C}$ of type $\mathsf{T}o$ ($o$ is an observation type), not of type $o$. Indeed, contexts should be allowed to do some computations: if they were of type $o$, they could only return values. In particular, a context $\mathbb{C}$ such that $x : \mathsf{T}\tau \vdash \mathbb{C} : o$ is derivable, meant to observe computations of type $\tau$, cannot observe anything, because the typing rule for the $\texttt{let}$ construct only allows us to use computations to build other computations, never values. Taking this into account, we get the following definition:

**Definition 1 (Contextual equivalence for $\lambda_{Comp}$).** *In $\lambda_{Comp}$, two values $a_1, a_2 \in [\![\tau]\!]$ are* contextually equivalent, *written as $a_1 \approx_\tau a_2$, if and only if, for all observable types $o \in \mathbf{Obs}$ and contexts $\mathbb{C}$ such that $x : \tau \vdash \mathbb{C} : \mathsf{T}o$ is derivable, $[\![\mathbb{C}]\!][x := a_1] = [\![\mathbb{C}]\!][x := a_2]$. Two closed terms $t_1$ and $t_2$ of type $\tau$ are contextually equivalent if and only if $[\![t_1]\!] \approx_\tau [\![t_2]\!]$. We use the same notation*

$\approx_\tau$ to denote the contextual equivalence for terms.

### 3.3 Logical relations for $\lambda_{Comp}$

A uniform framework for defining logical relations relies on the categorical notion of subscones [9], and a natural extension of logical relations able to deal with monadic types was introduced in [2]. The construction consists in lifting the CCC structure and the strong monad from the categorical model to the subscone. We reformulate this construction in the category $\mathcal{S}et$. The subscone is the category whose objects are binary relations $(A, B, R \subseteq A \times B)$ where $A$ and $B$ are sets; and a morphism between two objects $(A, B, R \subseteq A \times B)$ and $(A', B', R' \subseteq A' \times B')$ is a pair of functions $(f : A \to A', g : B \to B')$ preserving relations, i.e. $a\ R\ b \Rightarrow f(a)\ R'\ g(b)$.

The lifting of the CCC structure gives rise to the standard logical relations given in Section 2.2 and the lifting of the strong monad will give rise to relations for monadic types. We write $\tilde{T}$ for the lifting of the strong monad $T$. Given a relation $R \subseteq A \times B$ and two computations $a \in TA$ and $b \in TB$, $(a, b) \in \tilde{T}(R)$ if and only if there exists a computation $c \in T(R)$ (i.e. $c$ computes pairs in $R$) such that $a = T\pi_1(c)$ and $b = T\pi_2(c)$. The standard definition of logical relation for the simply typed $\lambda$-calculus is then extended with:

$$(c_1, c_2) \in \mathcal{R}_{\mathsf{T}\tau} \iff (c_1, c_2) \in \tilde{T}(\mathcal{R}_\tau). \tag{7}$$

This construction guarantees that Basic Lemma always holds provided that every constant is related to itself [2]. A list of instantiations of the above definition in concrete monads is also given in [2]. Figure 2 cites the relations for those monads defined in Figure 1.

| | |
|---|---|
| Partial computation: | $c_1\ \mathcal{R}_{\mathsf{T}\tau}\ c_2 \Leftrightarrow c_1\ \mathcal{R}_\tau\ c_2$ or $c_1 = c_2 = \bot$ |
| Exception: | $c_1\ \mathcal{R}_{\mathsf{T}\tau}\ c_2 \Leftrightarrow c_1\ \mathcal{R}_\tau\ c_2$ or $c_1 = c_2 \in E$ where $E$ is the set of exceptions |
| State transformer: | $c_1\ \mathcal{R}_{\mathsf{T}\tau}\ c_2 \Leftrightarrow \forall s \in St.\ \pi_1(c_1 s)\ \mathcal{R}_\tau\ \pi_1(c_2 s)\ \&\ \pi_2(c_1 s) = \pi_2(c_2 s)$ where $St$ is the set of states |
| Continuation: | $c_1\ \mathcal{R}_{\mathsf{T}\tau}\ c_2 \Leftrightarrow c_1(k_1) = c_2(k_2)$ for every $k_1, k_2$ such that $\forall a_1, a_2.\ a_1\ \mathcal{R}_\tau\ a_2 \implies k_1(a_1) = k_2(a_2)$ |
| Non-determinism: | $c_1\ \mathcal{R}_{\mathsf{T}\tau}\ c_2 \Leftrightarrow (\forall a_1 \in c_1.\ \exists a_2 \in c_2.\ a_1\ \mathcal{R}_\tau\ a_2)\ \&\ (\forall a_2 \in c_2.\ \exists a_1 \in c_1.\ a_1\ \mathcal{R}_\tau\ a_2)$ |

**Fig. 2.** Logical relations for concrete monads

We restrict our attention to logical relations $(\mathcal{R}_\tau)_{\tau \text{ type}}$ such that, for any observation type $o \in \mathbf{Obs}$, $\mathcal{R}_{\mathsf{T}o}$ is a partial equality. Such relations are called *observational* in the rest of the paper.

Note that we require partial identity on $\mathsf{T}o$, not on $o$. But if we assume that denotation of `val(_)`, i.e., the unit operation $\eta$, is injective, then that $\mathcal{R}_{\mathsf{T}o}$ is a partial equality implies that $\mathcal{R}_o$ is a partial equality as well. Indeed, let $a_1\ \mathcal{R}_o\ a_2$, and by Basic Lemma, $[\![\mathtt{val}(x)]\!][x := a_1]\ \mathcal{R}_{\mathsf{T}o}\ [\![\mathtt{val}(x)]\!][x := a_2]$, that is to say $\eta_{[\![o]\!]}(a_1) = \eta_{[\![o]\!]}(a_2)$. By injectivity of $\eta$, $a_1 = a_2$.

**Theorem 1 (Soundness of logical relations in $\lambda_{Comp}$).** *If $(\mathcal{R}_\tau)_{\tau \text{ type}}$ is an observational logical relation, then $\mathcal{R}_\tau \subseteq \approx_\tau$ for every type $\tau$.*

It is straightforward from the Basic Lemma.

### 3.4 Toward a proof on completeness of logical relations for $\boldsymbol{\lambda_{Comp}}$

Completeness of logical relations for $\lambda_{Comp}$ is much subtler than in $\lambda^{\rightarrow}$ due to the introduction of monadic types. We were expecting to find a general proof following the general construction defined in [2]. However, this turns out extremely difficult although it might not be impossible with certain restrictions, on types for example. The difficulty arises mainly from the different semantics for different forms of computations, which actually do not ensure that equivalent programs in one monad are necessarily equivalent in another. For instance, consider the following two programs in $\lambda_{Comp}$:

$$\texttt{let}\ x \Leftarrow t_1\ \texttt{in}\ \texttt{let}\ y \Leftarrow t_2\ \texttt{in}\ \texttt{val}(x),$$
$$\texttt{let}\ y \Leftarrow t_2\ \texttt{in}\ \texttt{let}\ x \Leftarrow t_1\ \texttt{in}\ \texttt{val}(x),$$

where both $t_1$ and $t_2$ are closed term. We can conclude that they are equivalent in the non-determinism monad — they return the same set of possible results of $t_1$, no matter what results $t_2$ produces, but this is not the case in, e.g., the exception monad when $t_1$ and $t_2$ throw different exceptions.

Being with such an obstacle, we shall switch our effort to case studies in Section 4 and we explore the completeness of logical relations for a list of common monads, precisely, all the monads listed in Figure 1. But, let us sketch out here a general structure for proving completeness of logical relations in $\lambda_{Comp}$. In particular, our study is still restricted to first-order types, which, in $\lambda_{Comp}$, are defined by the following grammar:

$$\tau^1 ::= b \mid \mathsf{T}\tau^1 \mid b \to \tau^1,$$

where $b$ ranges over the set of base types.

Similarly as in Proposition 1 in Section 2.2, we investigate completeness in a strong sense: we aim at finding an observational logical relation $(\mathcal{R}_\tau)_{\tau \text{ type}}$ such that if $\vdash t_1 : \tau$ and $\vdash t_2 : \tau$ are derivable and $t_1 \approx_\tau t_2$, for any type $\tau$ up to first order, then $[\![t_1]\!]\ \mathcal{R}_\tau\ [\![t_2]\!]$. Or briefly, $\sim_\tau \subseteq \mathcal{R}_\tau$, where $\sim_\tau$ is the relation defined in Section 2. As in the proof of Proposition 1, the logical relation $(\mathcal{R}_\tau)_{\tau \text{ type}}$ will be induced by $\mathcal{R}_b = \sim_b$, for any base type $b$. Then how to prove the completeness for an arbitrary monad $T$?

Note that we should also check that the logical relation $(\mathcal{R}_\tau)_{\tau \text{ type}}$, induced by $\mathcal{R}_b = \sim_b$, is observational, i.e., a partial equality on $\mathsf{T}o$, for any observable type $o$. Consider any pair $(a, b) \in \mathcal{R}_{\mathsf{T}o} = \tilde{T}(\mathcal{R}_o)$. By definition of the lifted monad $\tilde{T}$, there exists a computation $c \in T\mathcal{R}_o$ such that $a = T\pi_1(c)$ and $b = T\pi_2(c)$. But $\mathcal{R}_o = \sim_o \subseteq \mathrm{id}_{[\![o]\!]}$, hence the two projections $\pi_1, \pi_2 : \mathcal{R}_o \to [\![o]\!]$ are the same function, $\pi_1 = \pi_2$, and consequently $a = T\pi_1(c) = T\pi_2(c) = b$. This proves that $\mathcal{R}_{\mathsf{T}o}$ is a partial equality.

As usual, the proof of completeness would go by induction over $\tau$, to show $\sim_\tau \subseteq \mathcal{R}_\tau$ for each first-order type $\tau$. Cases $\tau = b$ and $\tau = b \to \tau'$ go identically as in $\lambda^{\rightarrow}$. The only difficult case is $\tau = \mathsf{T}\tau'$, i.e., the induction step:

$$\sim_\tau \subseteq \mathcal{R}_\tau \implies \sim_{\mathsf{T}\tau} \subseteq \mathcal{R}_{\mathsf{T}\tau} \qquad (8)$$

We did not find any general way to show (8) for an arbitrary monad. Instead, in the next section we prove it by cases, for all the monads in Figure 1 except the non-determinism monad. The non-determinism monad is an exceptional case where we do not have completeness for all first-order types but a subset of them. This will be studied separately in Section 4.3.

At the heart of the difficulty of showing (8), we find an issue of definability at monadic types in the set-theoretical model. We write $\mathsf{def}_\tau$ for the subset of definable elements in $[\![\tau]\!]$, and we eventually show that the relation between $\mathsf{def}_{\mathsf{T}\tau}$ and $\mathsf{def}_\tau$ can be shortly spelled-out:

$$\mathsf{def}_{\mathsf{T}\tau} \subseteq T\mathsf{def}_\tau \tag{9}$$

for all the monads we consider in this paper. This is a crucial argument for proving completeness of logical relations for monadic types, but to show (9), we need different proofs for different monads. This is detailed in Section 4.1.

## 4 Completeness of logical relations for monadic types

### 4.1 Definability in the set-theoretical model of $\lambda_{Comp}^{PESCN}$

As we have seen in $\lambda^{\to}$, definability is involved largely in the proof of completeness of logical relations (for first-order types). This is also the case in $\lambda_{Comp}$ and it apparently needs more concern due to the introduction of monadic types.

Despite we did not find a general proof for (9), it does hold for all the concrete monads in $\lambda_{Comp}^{PESCN}$. To state it formally, let us first define a predicate $\mathcal{P}_\tau$ on elements of $[\![\tau]\!]$, by induction on types:

- $\mathcal{P}_b = \mathsf{def}_b$, for every base type $b$;
- $\mathcal{P}_{\mathsf{T}\tau} = T(\mathsf{def}_\tau \cap \mathcal{P}_\tau)$;
- $\mathcal{P}_{\tau\to\tau'} = \{f \in \mathcal{P}_{\tau\to\tau'} \mid \forall a \in \mathsf{def}_\tau, f(a) \in \mathcal{P}_{\tau'}\}$.

We say that a constant c (of type $\tau$) is logical if and only if $\tau$ is a base type or $[\![c]\!] \in P_\tau$. We then require that $\lambda_{Comp}^{PESCN}$ contains only logical constants. Note that this restriction is valid because the predicates $P_{T\tau}$ and $P_{\tau\to\tau'}$ depend only on definability at type $\tau$. Some typical logical constants for monads in $\lambda_{Comp}^{PESCN}$ are as follows:

- Partial computation: a constant $\Omega_\tau$ of type $\mathsf{T}\tau$, for every $\tau$. $\Omega_\tau$ denotes the non-termination, so $[\![\Omega_\tau]\!] = \bot$.
- Exception: a constant $\mathtt{raise}_\tau^e$ of type $\mathsf{T}\tau$ for every type $\tau$ and every exception $e \in E$. $\mathtt{raise}_\tau^e$ does nothing but raises the exception $e$, so $[\![\mathtt{raise}_\tau^e]\!] = e$.
- State transformer: a constant $\mathtt{update}_s$ of type $\mathsf{T}\mathsf{unit}$ for every state $s \in St$, where $\mathsf{unit}$ is the base type which contains only a dummy value $*$. $\mathtt{update}_s$ simply changes the current state to $s$, so for any $s' \in St$, $[\![\mathtt{update}_s]\!](s') = (*, s)$.
- Continuation: a constant $\mathtt{call}_\tau^k$ of type $\tau \to T\mathsf{bool}$ for every $\tau$ and every continuation $k \in R^{[\![\tau]\!]}$. $\mathtt{call}_\tau^k$ calls directly the continuation $k$ — it behaves somehow like "goto" command, so for any $a \in [\![\tau]\!]$ and any continuation $k' \in R^{\mathsf{bool}}$, $[\![\mathtt{call}_\tau^k]\!](a)(k') = k(a)$.

– Non-determinism: a constant $+_\tau$ of type $\tau \to \tau \to \mathsf{T}\tau$ for every non-monadic type $\tau$. $+_\tau$ takes two arguments and returns randomly one of them — it introduces the non-determinism, so for any $a_1, a_2 \in [\![\tau]\!]$, $[\![+_\tau]\!](a_1, a_2) = \{a_1, a_2\}$.

We assume in the rest of this paper that the above constants are present in $\lambda_{Comp}^{PESCN}$. [1]

Note that $\mathcal{P}_\tau$ being a predicate on elements of $[\![\tau]\!]$ is equivalent to say that $\mathcal{P}_\tau$ can be seen as subset of $[\![\tau]\!]$, but in the case of monadic types, $\mathcal{P}_{\mathsf{T}\tau}$ (i.e., $T(\mathsf{def}_\tau \cap \mathcal{P}_\tau)$) is not necessary a subset of $[\![\mathsf{T}\tau]\!]$ (i.e., $T[\![\tau]\!]$). Fortunately, we prove that all the monads in $\lambda_{Comp}^{PESCN}$ preserves inclusions, which ensures that the predicate $\mathcal{P}$ is well-defined:

**Proposition 2.** *All the monads in $\lambda_{Comp}^{PESCN}$ preserve inclusions: $A \subseteq B \Rightarrow TA \subseteq TB$.*

*Proof.* We check it for every monad in $\lambda_{Comp}^{PESCN}$:

– Partial computation: according to the monad definition, if $A \subseteq B$, then for every $c \in TA$:

$$c \in TA \iff c \in A \text{ or } c = \bot \implies c \in B \text{ or } c = \bot \iff c \in TB.$$

– Exception: for every element $c \in TA$:

$$c \in TA \iff c \in A \text{ or } c \in E \implies c \in B \text{ or } c \in E \iff c \in TB.$$

– State transformer: for every $a \in TA$:

$$c \in TA \iff \forall s \in St \;.\; \pi_1(cs) \in A \implies \forall s \in St \;.\; \pi_1(cs) \in B \iff c \in TB.$$

– Continuation: this is a special case because apparently $TA = R^{R^A}$ is not a subset of $TB = R^{R^B}$, since they contain functions that are defined on different domains, but we shall consider here the functions coinciding on the smaller set $A$ as equivalent. We say that two functions $f_1$ and $f_2$ defined on a domain $B$ coincide on $A$ ($A \subseteq B$), written as $f_1|_A = f_2|_A$, if and only if for every $x \in A$, $f_1(x) = f_2(x)$. Then for every $c \in TA$:

$$\forall k_1, k_2 \in R^B \;.\; k_1 = k_2 \implies k_1|_A = k_2|_A \implies c(k_1) = c(k_2),$$

so $c$ is also function from $R^B$ to $R$, i.e., $c \in TB$.
– Non-determinism: for every $c \in TA$:

$$c \in TA \iff \forall a \in c \;.\; a \in A \implies \forall a \in c \;.\; a \in B \iff c \in TB. \qquad \square$$

Introducing such a constraint on constants is mainly for proving (9). Let us figure out the proof. Take an arbitrary element $c$ in $\mathsf{def}_{\mathsf{T}\tau}$. By definition, there exists a closed term $t$ of type $\mathsf{T}\tau$ such that $[\![t]\!] = c$. While it is not evident that $c \in T\mathsf{def}_\tau$, we are expecting to show that $[\![t]\!] \in T\mathsf{def}_\tau$, by considering the $\beta c$-normal-$\eta$-long form of $t$, since

[1] It is easy to check that each of these constants is related to itself, except $\mathtt{call}_\tau^k$ for continuations. However, we still assume the presence of $\mathtt{call}_\tau^k$ for the sake of proving completeness, while we are not able to prove the soundness with it. Note that Theorem 1 and Theorem 2 still hold, but they are not speaking of the same language.

$\lambda_{Comp}$ is strongly normalizing, Take the partial computation monad as an example, where $T\mathsf{def}_\tau = \mathsf{def}_\tau \cup \{\bot\}$. Consider the $\beta c$-normal-$\eta$-long form of $t$:

$$\mathtt{let}\; x_1 \Leftarrow d_1 u_{11} \cdots u_{1k_1}\; \mathtt{in} \cdots \mathtt{let}\; x_n \Leftarrow d_n u_{n1} \cdots u_{nk_n}\; \mathtt{in}\; \mathtt{val}(u).$$

We shall make the induction on $n$. It is clear that $[\![t]\!] \in T\mathsf{def}_\tau$ when $n = 0$. For the induction step, we hope that the closed term $d_1 u_{11} \cdots u_{1k_1}$ (of type $\mathsf{T}\tau_1$) would produce either $\bot$ (the non-termination), or a definable result (of type $\tau_1$) so that we can substitute $x_1$ in the rest of the normal term with the result of $d_1 u_{11} \cdots u_{1k_1}$ and make use of induction hypothesis. The constraint on constants helps here: to ensure that after the substitution, the resulted term is still in the proper form so that the induction would go through.

The following lemma shows that for every computation term $t$, $[\![t]\!] \in T\mathsf{def}_\tau$ if $t$ is in a particular form, which is a more general form of $\beta c$-normal-$\eta$-long form.

**Lemma 2.** *In $\lambda_{Comp}^{PESCN}$, $[\![t]\!] \in T\mathsf{def}_\tau$, for every closed computation term $t$ (of type $\mathsf{T}\tau$) of the following form:*

$$t \equiv \mathtt{let}\; x_1 \Leftarrow t_1 w_{11} \cdots w_{1k_1}\; \mathtt{in} \cdots \mathtt{let}\; x_n \Leftarrow t_n w_{n1} \cdots w_{nk_n}\; \mathtt{in}\; \mathtt{val}(w),$$

*where $n = 0, 1, 2, \ldots$ and $t_i$ ($1 \le i \le n$) is either a variable or a closed term such that $\mathcal{P}([\![t_i]\!])$ holds, and $w, w_{ij}$ ($1 \le i \le n, 1 \le j \le k_i$) are valid $\lambda_{Comp}^{PESCN}$ terms.*

*Proof.* We prove it by induction on $n$, for every monad:

– Partial computation ($T\mathsf{def}_\tau = \mathsf{def}_\tau \cup \{\bot\}$): if $n = 0$, it is clear that $[\![t]\!] \in T\mathsf{def}_\tau$. When $n > 0$, because $\mathcal{P}([\![t_1]\!])$ holds ($t_1$ must be closed), $[\![t_1 w_{11} \cdots w_{1k_1}]\!] \in T(\mathsf{def}_{\tau_1} \cap \mathcal{P}_{\tau_1})$. If $[\![t_1 w_{11} \cdots w_{1k_1}]\!] = \bot$, then $[\![t]\!] = \bot \in T\mathsf{def}_\tau$; otherwise, assume that $[\![t'_1]\!] = [\![t_1 w_{11} \cdots w_{1k_1}]\!]$ where $t'_1$ is a closed term of type $\tau_1$ (assuming that $t_1 w_{11} \cdots w_{1k_1}$ is of type $\mathsf{T}\tau_1$). According to the definition of $\mathcal{P}$, $\mathcal{P}([\![t'_1]\!])$ holds. Let $t'$ be another closed term:

$$t' \equiv \mathtt{let}\; x_2 \Leftarrow t'_2 w'_{21} \cdots w'_{2k_2}\; \mathtt{in} \cdots \mathtt{let}\; x_n \Leftarrow t'_n w'_{n1} \cdots w'_{nk_n}\; \mathtt{in}\; \mathtt{val}(w[t'_1/x_1]),$$

where $t'_i$ ($2 \le i \le n$) is either $t'_1$ or $t_i$, $w'_{ij} \equiv w_{ij}[t'_1/x_1]$ ($2 \le i \le n, 1 \le j \le k_i$). By induction, $[\![t']\!] \in T\mathsf{def}_\tau$ holds. Furthermore,

$$\begin{aligned}[\![t']\!] &= [\![\mathtt{let}\; x_2 \Leftarrow t_2 w_{21} \cdots w_{2k_2}\; \mathtt{in} \cdots \\ &\qquad \mathtt{let}\; x_n \Leftarrow t_n w_{n1} \cdots w_{nk_n}\; \mathtt{in}\; \mathtt{val}(w)]\!][x_1 := [\![t'_1]\!]] \\ &= [\![\mathtt{let}\; x_1 \Leftarrow t_1 w_{11} \cdots w_{1k_1}\; \mathtt{in} \cdots \mathtt{let}\; x_n \Leftarrow t_n w_{n1} \cdots w_{nk_n}\; \mathtt{in}\; \mathtt{val}(w)]\!] \\ &= [\![t]\!],\end{aligned}$$

hence $[\![t]\!] \in T\mathsf{def}_\tau$.

– Exception ($T\mathsf{def}_\tau = \mathsf{def}_\tau \cup E$): if $n = 0$, clearly $[\![t]\!] \in T\mathsf{def}_\tau$. When $n > 0$, because $\mathcal{P}([\![t_1]\!])$ holds, $[\![t_1 w_{11} \cdots w_{1k_1}]\!] \in T(\mathsf{def}_{\tau_1} \cap \mathcal{P}_{\tau_1})$. If $[\![t_1 w_{11} \cdots w_{1k_1}]\!] \in E$, then $[\![t]\!] \in E \subseteq T\mathsf{def}_\tau$; otherwise, exactly as in the case of partial computation, build a term $t'$. Similarly, we prove that $[\![t]\!] = [\![t']\!] \in T\mathsf{def}_\tau$ by induction.

– State transformer ($T\mathsf{def}_\tau = (\mathsf{def}_\tau \times St)^{St}$): when $n = 0$, for every $s \in St$, $\pi^1([\![t]\!]s) = [\![w]\!] \in \mathsf{def}_\tau$ hence $[\![t]\!] \in T\mathsf{def}_\tau$. When $n > 0$, for every $s \in St$, assume that $[\![t_1^s]\!] = \pi^1([\![t_1 w_{11} \cdots w_{1k_1}]\!]s)$ where $t_1'$ is a closed term of type $\tau_1$ (assuming that $t_1 w_{11} \cdots w_{1k_1}$ is of type $\mathsf{T}\tau_1$). According to the definition of $\mathcal{P}$, $\mathcal{P}([\![t_1^s]\!])$ holds. Let $t^s$ be another closed term:

$$t^s \equiv \texttt{let}\ x_2 \Leftarrow t_2^s w_{21}^s \cdots w_{2k_2}^s\ \texttt{in} \cdots \texttt{let}\ x_n \Leftarrow t_n^s w_{n1}^s \cdots w_{nk_n}^s\ \texttt{in}\ \texttt{val}(w[t_1^s/x_1]),$$

where $t_i^s$ $(2 \le i \le n)$ is either $t_1^s$ or $t_i$, $w_{ij}^s \equiv w_{ij}[t_1^s/x_1]$ $(2 \le i \le n, 1 \le j \le k_i)$. By induction, $[\![t^s]\!] \in T\mathsf{def}_\tau$ holds. Furthermore, for every $s \in St$,

$$\begin{aligned}[\![t]\!]s &= [\![\texttt{let}\ x_1 \Leftarrow t_1 w_{11} \cdots w_{1k_1}\ \texttt{in} \cdots \texttt{let}\ x_n \Leftarrow t_n w_{n1} \cdots w_{nk_n}\ \texttt{in}\ \texttt{val}(w)]\!]s \\ &= ([\![\texttt{let}\ x_2 \Leftarrow t_2 w_{21} \cdots w_{2k_2}\ \texttt{in} \cdots \\ &\qquad \texttt{let}\ x_n \Leftarrow t_n w_{n1} \cdots w_{nk_n}\ \texttt{in}\ \texttt{val}(w)]\!][x_1 := [\![t_1^s]\!]])s' \\ &= [\![t^s]\!]s',\end{aligned}$$

where $s' = \pi_2([\![t_1 w_{11} \cdots w_{1k_1}]\!]s)$. Since $[\![t^s]\!] \in T\mathsf{def}_\tau$ for every $s \in St$, $\pi_1([\![t]\!]s) = \pi_1([\![t^s]\!]s') \in \mathsf{def}_\tau$, hence $[\![t]\!] \in T\mathsf{def}_\tau$.

– Continuation ($T\mathsf{def}_\tau = R^{R^{\mathsf{def}_\tau}}$): we say that an element $c \in [\![\mathsf{T}\tau]\!] = R^{R^{[\![\tau]\!]}}$ is in $T\mathsf{def}_\tau$ if and only if for every pair of continuations $k_1, k_2 \in R^{[\![\tau]\!]}$,

$$k_1|_{\mathsf{def}_\tau} = k_2|_{\mathsf{def}_\tau} \Longrightarrow c(k_1) = c(k_2).$$

If $n = 0$, $[\![t]\!] = \underline{\lambda} k.k([\![w]\!]) \in T\mathsf{def}_\tau$. When $n > 0$, according to the definition of the continuation monad: $[\![t]\!] = \underline{\lambda} k \cdot [\![t_1 w_{11} \cdots w_{nk_n}]\!](k')$, where

$$k' = \underline{\lambda} a \cdot ([\![\texttt{let}\ x_2 \Leftarrow t_2 w_{21} \cdots w_{2k_2}\ \texttt{in} \cdots \texttt{let}\ x_n \Leftarrow t_n w_{n1} \cdots w_{nk_n}\ \texttt{in}\ \texttt{val}(w)]\!][x_1 := a])k.$$

For every continuations $k_1, k_2 \in R^{[\![\tau]\!]}$ such that $k_1|_{\mathsf{def}_\tau} = k_2|_{\mathsf{def}_\tau}$ let

$$k_i' = \underline{\lambda} a \cdot ([\![\texttt{let}\ x_2 \Leftarrow t_2 w_{21} \cdots w_{2k_2}\ \texttt{in} \cdots \texttt{let}\ x_n \Leftarrow t_n w_{n1} \cdots w_{nk_n}\ \texttt{in}\ \texttt{val}(w)]\!][x_1 := a])k_i,$$

$i = 1, 2$. Because $[\![t_1 w_{11} \cdots w_{1k_1}]\!] \in T(\mathcal{P}_{\tau_1} \cap \mathsf{def}_{\tau_1})$, if we can prove $k_1'|_{\mathcal{P}_{\tau_1} \cap \mathsf{def}_{\tau_1}} = k_2'|_{\mathcal{P}_{\tau_1} \cap \mathsf{def}_{\tau_1}}$, which implies $[\![t]\!](k_1) = [\![t]\!](k_2)$, we can conclude $[\![t]\!] \in T\mathsf{def}_\tau$. For every $a \in \mathcal{P}_{\tau_1} \cap \mathsf{def}_{\tau_1}$, let $[\![t_1^a]\!] = a$ where $t_1^a$ is a closed term. Define another closed term $t^a$:

$$t^a \equiv \texttt{let}\ x_2 \Leftarrow t_2^a w_{21}^a \cdots w_{2k_2}^a\ \texttt{in} \cdots \texttt{let}\ x_n \Leftarrow t_n^a w_{n1}^a \cdots w_{nk_n}^a\ \texttt{in}\ \texttt{val}(w[t_1^a/x_1]),$$

where $t_i^a$ $(2 \le i \le n)$ is either $t_1^a$ or $t_i$, $w_{ij}^a \equiv w_{ij}[t_1^a/x_1]$ $(2 \le i \le n, 1 \le j \le k_i)$. By induction, $[\![t^a]\!] \in T\mathsf{def}_\tau$, so $k_1'(a) = [\![t^a]\!]k_1 = [\![t^a]\!]k_2 = k_2'(a)$, i.e., $k_1'|_{\mathcal{P}_{\tau_1} \cap \mathsf{def}_{\tau_1}} = k_2'|_{\mathcal{P}_{\tau_1} \cap \mathsf{def}_{\tau_1}}$.

– Non-determinism ($T\mathsf{def}_\tau = \mathbb{P}_{\mathrm{fin}}(\mathsf{def}_\tau)$): when $n = 0$, $[\![t]\!] = \{[\![w]\!]\} \in T\mathsf{def}_\tau$. When $n > 0$, for every $a \in [\![t_1 w_{11} \cdots w_{1k_1}]\!]$, assume that $[\![t_1^a]\!] = a$ where $t_1'$ is a closed term of type $\tau_1$ (assuming that $t_1 w_{11} \cdots w_{1k_1}$ is of type $\mathsf{T}\tau_1$). According to the definition of $\mathcal{P}$, $\mathcal{P}([\![t_1^a]\!])$ holds. Let $t^a$ be another closed term:

$$t^a \equiv \texttt{let}\ x_2 \Leftarrow t_2^a w_{21}^a \cdots w_{2k_2}^a\ \texttt{in} \cdots \texttt{let}\ x_n \Leftarrow t_n^a w_{n1}^a \cdots w_{nk_n}^a\ \texttt{in}\ \texttt{val}(w[t_1^a/x_1]),$$

where $t_i^a$ $(2 \le i \le n)$ is either $t_1^a$ or $t_i$, $w_{ij}^a \equiv w_{ij}[t_1^a/x_1]$ $(2 \le i \le n, 1 \le j \le k_i)$. By induction, $[\![t^a]\!] \in T\mathsf{def}_\tau$ holds. Furthermore,

$$\begin{aligned}[\![t]\!] &= [\![\mathtt{let}\ x_1 \Leftarrow t_1 w_{11} \cdots w_{1k_1}\ \mathtt{in} \cdots \mathtt{let}\ x_n \Leftarrow t_n w_{n1} \cdots w_{nk_n}\ \mathtt{in}\ \mathtt{val}(w)]\!] \\ &= \bigcup_{a \in [\![t_1]\!]} [\![\mathtt{let}\ x_2 \Leftarrow t_2 w_{21} \cdots w_{2k_2}\ \mathtt{in} \cdots \\ &\qquad \mathtt{let}\ x_n \Leftarrow t_n w_{n1} \cdots w_{nk_n}\ \mathtt{in}\ \mathtt{val}(w)]\!][x_1 := a] \\ &= \bigcup_{a \in [\![t_1]\!]} [\![t^a]\!].\end{aligned}$$

Because $[\![t^a]\!] \in T\mathsf{def}_\tau$ holds for every $a \in [\![t_1 w_{11} \cdots w_{1k_1}]\!]$, $[\![t]\!] \in T\mathsf{def}_\tau$. □

From the above lemma, we conclude immediately that for every closed $\beta c$-normal-$\eta$-long computation term $t$ in $\lambda_{Comp}^{PESCN}$ with logical constants, $[\![t]\!] \subseteq T\mathsf{def}_\tau$.

**Proposition 3.** $\mathsf{def}_{\mathsf{T}\tau} \subseteq T\mathsf{def}_\tau$ *holds in the set-theoretical model of* $\lambda_{Comp}^{PESCN}$ *with logical constants.*

*Proof.* It follows from Lemma 2 by considering the $\beta c$-normal-$\eta$-long terms that define elements in $[\![\mathsf{T}\tau]\!]$ since $\lambda_{Comp}$ is strongly normalizing. □

### 4.2 Completeness of logical relations in $\lambda_{Comp}^{PESC}$ for first-order types

We prove (8) in this section for the partial computation monad, the exception monad, the state monad and the continuation monad. We write $\lambda_{Comp}^{PESC}$ for $\lambda_{Comp}$ where the monad is restricted to one of these four monads.

Proofs depend typically on the particular semantics of every form of computation, but a common technique is used frequently: given two definable but non-related elements of $[\![\mathsf{T}\tau]\!]$, one can find a context to distinguish the programs (of type $\mathsf{T}\tau$) that define the two given elements, and such a context is usually built based on another context that can distinguish programs of type $\tau$.

**Lemma 3.** *Let* $(\mathcal{R}_\tau)_{\tau \text{ type}}$ *be a logical relation in* $\lambda_{Comp}^{PESC}$ *with only logical constants.* $\sim_\tau \subseteq \mathcal{R}_\tau \implies \sim_{\mathsf{T}\tau} \subseteq \mathcal{R}_{\mathsf{T}\tau}$ *holds for every type* $\tau$.

*Proof.* Take two arbitrary elements $c_1, c_2 \in [\![\mathsf{T}\tau]\!]$ such that $(c_1, c_2) \notin \mathcal{R}_{\mathsf{T}\tau}$, we prove that $c_1 \not\sim_{\mathsf{T}\tau} c_2$ for every monad in $\lambda_{Comp}^{PESC}$:

- Partial computation: the fact $(c_1, c_2) \notin \mathcal{R}_{\mathsf{T}\tau}$ amounts to the following two cases:
  - $c_1, c_2 \in [\![\tau]\!]$ but $(c_1, c_2) \notin \mathcal{R}_\tau$, then $c_1 \not\sim_\tau c_2$. If one of these two values is not definable at type $\tau$, by Proposition 3, it is not definable at type $\mathsf{T}\tau$ either. If both values are definable at type $\tau$ but they are not contextually equivalent, then there is a context $x : \tau \vdash \mathbb{C} : \mathsf{T}o$ such that $[\![\mathbb{C}]\!][x := c_1] \neq [\![\mathbb{C}]\!][x := c_2]$. Thus, the context $y : \mathsf{T}\tau \vdash \mathtt{let}\ x \Leftarrow y\ \mathtt{in}\ \mathbb{C} : \mathsf{T}o$ can distinguish $c_1$ and $c_2$ (as two values of type $\mathsf{T}\tau$).
  - $c_1 \in [\![\tau]\!]$ and $c_2 = \bot$ (or symmetrically, $c_1 = \bot$ and $c_2 \in [\![\tau]\!]$), then the context $\mathtt{let}\ x \Leftarrow y\ \mathtt{in}\ \mathtt{val}(\mathtt{true})$ can be used to distinguish them.

  $c_1 \not\sim_{\mathsf{T}\tau} c_2$ in both cases.
- Exception: the fact $(c_1, c_2) \notin \mathcal{R}_{\mathsf{T}\tau}$ amounts to three cases:

- $c_1, c_2 \in [\![\tau]\!]$ but $(c_1, c_2) \notin \mathcal{R}_\tau$, then $c_1 \not\sim_\tau c_2$. Suppose both values are definable at type $\tau$, otherwise by Proposition 3, they must not be definable at type $\mathsf{T}\tau$. Similar as in the case of partial computation we can build a context that distinguishes $c_1$ and $c_2$ as values of type $\mathsf{T}\tau$, from the context that distinguishes $c_1$ and $c_2$ as values of type $\tau$.
- $c_1 \in [\![\tau]\!], c_2 \in E$. Consider the following context:

$$y : \mathsf{T}\tau \vdash \texttt{let } x \Leftarrow y \texttt{ in val}(\texttt{true}) : \mathsf{T}\mathsf{bool}.$$

  When $y$ is substituted by $c_1$ and $c_2$, the context evaluates to different values, namely, a boolean and an exception.
- $c_1, c_2 \in E$ but $c_1 \neq c_2$. Try the same context as in the second case, which will evaluate to two different exceptions that can be distinguished.

$c_1 \not\sim_{\mathsf{T}\tau} c_2$ in all the three cases.

– State transformer: because $(c_1, c_2) \notin \mathcal{R}_{\mathsf{T}\tau}$, there exists some $s_0 \in St$ such that
  - either $(\pi_1(c_1 s_0), \pi_1(c_2 s_0)) \notin \mathcal{R}_\tau$. Then by induction $\pi_1(c_1 s_0) \not\sim_\tau \pi_1(c_2 s_0)$. If $\pi_1(c_i s_0)$ $(i = 1, 2)$ is not definable, then by Proposition 3, $c_i$ is not definable either. If both $\pi_1(c_1 s_0)$ and $\pi_1(c_2 s_0)$ are definable, but $\pi_1(c_1 s_0) \not\approx_\tau \pi_1(c_2 s_0)$, then there is a context $x : \tau \vdash \mathbb{C} : \mathsf{T}o$ such that $[\![\mathbb{C}]\!][x := \pi_1(c_1 s_0)] \neq [\![\mathbb{C}]\!][x := \pi_1(c_2 s_0)]$, i.e., for some state $s'_0 \in St$,

$$[\![\mathbb{C}]\!][x := \pi_1(c_1 s_0)](s'_0) \neq [\![\mathbb{C}]\!][x := \pi_1(c_1 s_0)](s'_0).$$

    Now we can use the following context:

$$y : \mathsf{T}\tau \vdash \texttt{let } x \Leftarrow y \texttt{ in let } z \Leftarrow \texttt{update}_{s'_0} \texttt{ in } \mathbb{C} : To,$$

    Let $f_i = \left[\!\!\left[\texttt{let } x \Leftarrow y \texttt{ in let } z \Leftarrow \texttt{update}_{s'_0} \texttt{ in } \mathbb{C}\right]\!\!\right][y := c_i]$, then for every $s \in St$,

$$\begin{aligned} f_i(s) &= \left[\!\!\left[\texttt{let } z \Leftarrow \texttt{update}_{s'_0} \texttt{ in } \mathbb{C}\right]\!\!\right][x := \pi_1(c_i s)](\pi_2(c_i s)) \\ &= [\![\mathbb{C}]\!][x := \pi_1(c_i s)](s'_0), \qquad (i = 1, 2). \end{aligned}$$

    $f_1 \neq f_2$, because when applied to the state $s_0$, they will return two different pairs, so the above context can distinguish the two values $c_1$ and $c_2$;
  - or $\pi_2(c_1 s_0) \neq \pi_2(c_2 s_0)$. we use the context

$$y : \mathsf{T}\tau \vdash \texttt{let } x \Leftarrow y \texttt{ in val}(\texttt{true}) : \mathsf{T}\mathsf{bool},$$

    then $[\![\texttt{let } x \Leftarrow y \texttt{ in val}(\texttt{true})]\!][y := c_i] = \lambda s.(\texttt{true}, \pi_2(c_i s))$ $(i = 1, 2)$. These two functions are not equal since they return different results when applied to the state $s_0$.

  In both cases, $c_1 \not\sim_{\mathsf{T}\tau} c_2$.

– Continuation: first say that two continuations $k_1, k_2 \in R^{[\![\tau]\!]}$ are *$\mathcal{R}$-related*, if and only if for every $a_1, a_2 \in [\![\tau]\!]$, $a_1 \mathcal{R}_\tau a_2 \Longrightarrow k_1(a_1) = k_2(a_2)$. The fact $(c_1, c_2) \notin$ $\mathcal{R}_{\mathsf{T}\tau}$ means that there are two $\mathcal{R}$-related continuations $k_1, k_2$ such that $c_1(k_1) \neq c_2(k_2)$. Because $\sim_\tau \subseteq \mathcal{R}_\tau$, for every definable value $a \in \mathsf{def}_\tau$, clearly,

$$a \sim_\tau a \Longrightarrow a_1 \ \mathcal{R} \ a_2 \Longrightarrow k_1(a_1) = k_2(a_2),$$

so $k_1$ and $k_2$ coincide over $\mathsf{def}_\tau$. Suppose that both $c_1$ and $c_2$ are definable, then by Proposition 3, $c_1(k_1) = c_1(k_2)$ and $c_2(k_1) = c_2(k_2)$, hence $c_1(k_1) \neq c_2(k_1)$. Consider the context

$$y : \mathsf{T}\tau \vdash \mathtt{let}\ x \Leftarrow y\ \mathtt{in}\ \mathtt{call}_\tau^{k_1}(x) : T\mathsf{bool}.$$

For every $k \in R^{[\![\mathsf{bool}]\!]}$,

$$\begin{aligned} & [\![\mathtt{let}\ x \Leftarrow y\ \mathtt{in}\ \mathtt{call}_\tau^{k_1}(x)]\!]\,[y := c_i](k) \qquad (i = 1, 2), \\ = {} & c_i(\underline{\lambda}\, a \cdot ([\![\mathtt{call}_\tau^{k_1}(x)]\!]\,[x := a])k) \\ = {} & c_i(\underline{\lambda}\, a \cdot k_1(a)) \ = \ c_i(k_1). \end{aligned}$$

Since $c_1(k_1) \neq c_2(k_1)$, this context distinguishes the two computations, hence $c_1 \not\sim_{\mathsf{T}\tau} c_2$. □

**Theorem 2.** *In $\lambda_{Comp}^{PESC}$, if all constants are logical and in particular, if the following constants are present*

– $\mathtt{update}_s$ *for the state transformer monad;*
– $\mathtt{call}_\tau^k$ *for the continuation monad,*

*then logical relations are complete up to first-order types, in the strong sense that there exists an observational logical relation $(\mathcal{R}_\tau)_{\tau\ type}$ such that for any closed terms $t_1, t_2$ of any type $\tau^1$ up to first order, if $t_1 \approx_{\tau^1} t_2$, then $[\![t_1]\!]\ \mathcal{R}_{\tau^1}\ [\![t_2]\!]$.*

*Proof.* Take the logical relation $(\mathcal{R}_\tau)_{\tau\ type}$ induced by $\mathcal{R}_b = \sim_b$, for any base type $b$. We prove by induction on types that $\sim_\tau \subseteq \mathcal{R}_\tau$ for any first-order type $\tau$. In particular, the induction step $\sim_\tau \subseteq \mathcal{R}_\tau \Longrightarrow \sim_{\mathsf{T}\tau}; \subseteq \mathcal{R}_{\mathsf{T}\tau}$ is shown by Lemma 3. □

### 4.3 Completeness of logical relations for the non-determinism monad

The non-determinism monad is an interesting case: the completeness of logical relations for this monad does not hold for all first-order types! To state it, consider the following two programs of a first-order type that break the completeness of logical relations:

$$\begin{aligned} & \vdash \mathtt{val}(\lambda x.(\mathtt{true} +_{\mathsf{bool}} \mathtt{false})) : \mathsf{T}(\mathsf{bool} \to \mathsf{T}\mathsf{bool}), \\ & \vdash \lambda x.\mathtt{val}(\mathtt{true})\ +_{\mathsf{bool}\to\mathsf{T}\mathsf{bool}}\ \lambda x.(\mathtt{true} +_{\mathsf{bool}} \mathtt{false}) : \mathsf{T}(\mathsf{bool} \to \mathsf{T}\mathsf{bool}). \end{aligned}$$

Recall the logical constant $+_\tau$ of type $\tau \to \tau \to \mathsf{T}\tau$: $[\![+_\tau]\!](a_1, a_2) = \{a_1, a_2\}$ for every $a_1, a_2 \in [\![\tau]\!]$. The two programs are contextually equivalent: what contexts can do is to apply the functions to some arguments and observe the results. But no matter how many time we apply these two functions, we always get the same set of possible values ($\{\texttt{true}, \texttt{false}\}$), so there is no way to distinguish them with a context. Recall the logical relation for non-determinism monad in Figure 2:

$$c_1 \; \mathcal{R}_{\mathsf{T}\tau} \; c_2 \Leftrightarrow (\forall a_1 \in c_1.\ \exists a_2 \in c_2.\ a_1 \; \mathcal{R}_\tau \; a_2) \;\; \& \;\; (\forall a_2 \in c_2.\ \exists a_1 \in c_1.\ a_1 \; \mathcal{R}_\tau \; a_2).$$

Clearly the denotations of the above two programs are not related by that relation because the function $[\![\lambda x.\texttt{val}(\texttt{true})]\!]$ from the second program is not related to the function in the first.

However, if we assume that for every non-observable base type $b$, there is an equality test constant $\texttt{test}_b : b \to b \to \mathsf{bool}$ (clearly, $\mathcal{P}(\texttt{test}_b)$ holds), logical relations for the non-determinism monad are then complete for a set of *weak first-order types*:

$$\tau^1_{\mathrm{w}} ::= b \mid \mathsf{T}b \mid b \to \tau^1_{\mathrm{w}}.$$

Compared to all types up to first order, weak first-order types do not contain monadic types of functions, so it immediately excludes the two programs in the above counterexample.

**Theorem 3.** *Logical relations for the non-determinism monad are complete up to weak first-order types. in the strong sense that there exists an observational logical relation $(\mathcal{R}_\tau)_{\tau \text{ type}}$ such that for any closed terms $t_1, t_2$ of a weak first-order type $\tau^1_w$, if $t_1 \approx_{\tau^1_w} t_2$, then $[\![t_1]\!] \; \mathcal{R}_{\tau^1_w} \; [\![t_2]\!]$.*

*Proof.* Take the logical relation $\mathcal{R}$ induced by $\mathcal{R}_b = \sim_b$, for any base type $b$. We prove by induction on types that $\sim_{\tau^1_w} \subseteq \mathcal{R}_{\tau^1_w}$ for any weak first-order type $\tau^1_w$.

Cases $b$ and $b \to \tau^1_w$ go identically as in standard typed lambda-calculi. For monadic types $\mathsf{T}b$, suppose that $(c_1, c_2) \notin \mathcal{R}_{\mathsf{T}b}$, which means either there is a value in $c_1$ such that no value of $c_2$ is related to it, or there is such a value in $c_2$. We assume that every value in $c_1$ and $c_2$ is definable (otherwise it is obvious that $c_1 \not\sim_{\mathsf{T}b} c_2$ because at least one of them is not definable, according to Proposition 3). Suppose there is a value $a \in c_1$ such that no value in $c_2$ is related to it, and $a$ can be defined by a closed term $t$ of type $b$. Then the following context can distinguish $c_1$ and $c_2$:

$$x : \mathsf{T}\tau \vdash \texttt{let}\ y \Leftarrow x\ \texttt{in}\ \texttt{test}_b(y, t) : \mathsf{Tbool}$$

since every value in $c_2$ is not contextually equivalent to $a$, hence not equal to $a$. □

Now let $\mathsf{state}$ and $\mathsf{label}$ be base types such that $\mathsf{label}$ is an observation type, whereas $\mathsf{state}$ is not. Using non-determinism monad, we can define labeled transition systems as elements of $[\![\mathsf{state} \to \mathsf{label} \to \mathsf{Tstate}]\!]$, with states in $[\![\mathsf{state}]\!]$ and labels in $[\![\mathsf{label}]\!]$, as functions mapping states $a$ and labels $l$ to the set of states $b$ such that $a \xrightarrow{l} b$ . The logical relation at type $\mathsf{state} \to \mathsf{label} \to \mathsf{Tstate}$ is given by [2]:

$$\begin{array}{l}
(f_1, f_2) \in \mathcal{R}_{\mathsf{state}\to\mathsf{label}\to\mathsf{Tstate}} \iff \\
\quad \forall a_1, a_2, l_1, l_2 \cdot (a_1, a_2) \in \mathcal{R}_{\mathsf{state}} \;\; \& \;\; (l_1, l_2) \in \mathcal{R}_{\mathsf{label}} \implies \\
\qquad (\forall b_1 \in f_1(a_1, l_1) \cdot \exists b_2 \in f_2(a_2, l_2) \cdot (b_1, b_2) \in \mathcal{R}_{\mathsf{state}}) \\
\qquad \;\; \& \;\; (\forall b_2 \in f_2(a_2, l_2) \cdot \exists b_1 \in f_1(a_1, l_1) \cdot (b_1, b_2) \in \mathcal{R}_{\mathsf{state}})
\end{array}$$

In case $\mathcal{R}_{\mathsf{label}}$ is equality, $f_1$ and $f_2$ are logically related if and only if $\mathcal{R}_{\mathsf{state}}$ is a *strong bisimulation* between the labeled transition systems $f_1$ and $f_2$.

Sometimes we explicitly specify an initial state for certain labeled transition system. In this case, the encoding of the labeled transition system in the nondeterminism monad is a pair $(q, f)$ of $[\![\mathsf{state} \times (\mathsf{state} \to \mathsf{label} \to \mathsf{Tstate})]\!]$, where $q$ is the initial state and $f$ is the transition relation as defined above. Then $(q_1, f_1)$ and $(q_2, f_2)$ are logically related if and only if they are strongly bisimular, i.e., $\mathcal{R}_{\mathsf{state}}$ is a strong bisimulation between the two labeled transition systems and $q_1 \mathcal{R}_{\mathsf{state}} q_2$.

**Corollary 1 (Soundness of strong bisimulation).** *Let $f_1$ and $f_2$ be transition systems. If there exists a strong bisimulation between $f_1$ and $f_2$, then $f_1$ and $f_2$ are contextually equivalent.*

*Proof.* There exists a strong bisimulation between $f_1$ and $f_2$, therefore $f_1$ and $f_2$ are logically related. By Theorem 1, $f_1$ and $f_2$ are thus contextually equivalent. □

In order to prove completeness, we need to assume that label has no *junk*, in the sense that every value of $[\![\mathsf{label}]\!]$ is definable.

**Corollary 2 (Completeness of strong bisimulation).** *Let $f_1$ and $f_2$ be transition systems which are definable. If $f_1$ and $f_2$ are contextually equivalent and* label *has no junk, then there exists a strong bisimulation between $f_1$ and $f_2$.*

*Proof.* Let $\mathcal{R}$ be the logical relation given by Theorem 3. $f_1$ and $f_2$ are definable and contextually equivalent, so $f_1 \ \mathcal{R}_{\mathsf{state} \to \mathsf{label} \to \mathsf{Tstate}} \ f_2$. Moreover, because label has no junk, $\mathcal{R}_{\mathsf{label}}$ is equality. $\mathcal{R}_{\mathsf{state}}$ is thus a strong bisimulation between $f_1$ and $f_2$. □

## 5 Conclusion

The work presented in this paper is a natural continuation of the authors' previous work [2,3]. In [2], we extend [9] and derive logical relations for monadic types which are sound in the sense that the Basic Lemma still holds. In [3], we study contextual equivalence in a specific version of the computational $\lambda$-calculus with cryptographic primitives and we show that lax logical relations (the categorical generalization of logical relations [14]) derived using the same construction is complete. Then in this paper, we explore the completeness of logical relations for the computational $\lambda$-calculus and we show that they are complete at first-order types, for a list of common monads: partial computations, exceptions, state transformers and continuations, while in the case of continuation, the completeness depends on a natural constant `call`, with which we cannot show the soundness.

Pitts and Stark have defined operationally based logical relations to characterize the contextual equivalence in a language with local store [13]. This work can be traced back to their early work on the nu-calculus [12] which can be translated in a special version of the computational $\lambda$-calculus and be modeled using the dynamic name creation monad [17]. Logical relations for this monad are derived in [19] using the construction from [2]. It is also shown in [19] that such derived logical relations are equivalent to Pitts and Stark's operational logical relations up to second-order types.

An exceptional case of our completeness result is the non-determinism monad, where logical relations are not complete for all first-order types, but a subset of them. We effectively show this by providing a counter-example that breaks the completeness at first-order types. This is indeed an interesting case. A more comprehensive study on this monad can be found in [4], where Jeffrey defines a denotational model for the computational $\lambda$-calculus specialized in non-determinism and proves that this model is fully abstract for may-testing. The relation between our notion of contextual equivalence and the may-testing equivalence remains to be clarified.

Recently, Lindley and Stark introduce the syntactic $\top\top$-lifting for the computational $\lambda$-calculus and prove the strong normalization [7]. Katsumata then instantiates their liftings in $\mathcal{S}et$ [5]. The $\top\top$-lifting of strong monads is an essentially different approach from that in [2]. It would be interesting to establish a formal relationship between these two approaches, and to look for a general proof of completeness using the $\top\top$-lifting.